# When none of us perform better than all of us together: the role of analogical decision rules in groups[i]

NICOLETA MESLEC, PETRU L. CURSEU, MARIUS T.H.MEEUS, Tilburg University
OANA C. FODOR (IEDERAN), Babes-Bolyai University

## 1. INTRODUCTION

Reliance on groups in social life is built on a strong assumption, namely that the array of information exchanged, explored and integrated in groups enhances decision quality relative to individual choices (Hinsz, 1990;). Similarly, other species organize and work in collectives in order to enhance their survival chances. For example, homing and migrating birds collectively decide on communal routes that maximize their chances of survival and successful arrival to their destination and swarms of bees and ants collectively choose new nest sites on which their survival depends (Conradt &List, 2009; Sasaki & Pratt, 2012; Edwards & Pratt, 2009). Social interactions unfolding in such collectives shape the emergence of collective choices that transcend a simple aggregation of individual preferences or competencies (Curseu & Schruijer, 2012; Krause, Ruxton & Krause, 2010).

Although groups have the potential to become superior (as interacting collectives) to standalone individuals, this (emergent) potential is not always realized in real-life situations. Studies stemming from the group synergy literature illustrate not only that groups do not manage to achieve strong cognitive synergy (perform better than their best individual member - Laughlin, Gonzalez & Sommer, 2003; Meslec & Curseu, 2013) but sometimes they even have difficulties to achieve weak cognitive synergy (they perform worse than the average individual performance in the group - Buehler, Messervey & Griffin, 2005; Hinsz, Tindale & Nagao, 2008). This paper investigates experimentally in two studies how decision rules (collaboration vs. identify-the-best) and the way in which are induced (direct vs. analogic) affect group synergy.

## 2. STUDY 1

The first study contrasts the collaborative rule with a heuristic rule, namely identify-the-best which is inspired from the ecological rationality view. The collaborative rule encourages opinion sharing and equal participation of all group members during deliberations. Although the collaborative rule increases the information processing efforts in groups, it also has shortcomings: (1) in absolute terms has not yet been proved to lead to strong cognitive synergy (Curseu, Jansen & Chappin, 2013), and (2) it comes with costs in terms of time and cognitive resources that need to be invested in the group decision. Identify-the-best heuristic requires group members to identify the most capable member in the group and to improve his/her performance. In line with ecological rationality, we argue that a decision rule such as identify-the-best is particularly relevant to cognitive synergy, given that the core of strong synergy lies in groups outperforming its best individual member.

Next to decision rule content we also manipulate the way in which decision rules are induced (direct vs. analogic). Recent experimental research only explored the effects of directly induced decision rules (Curseu & Schruijer, 2012; Curseu et al., 2013). Nevertheless, decision rules with the potential to foster strong synergy may stem from analogies made with successful groups in the environment. Via analogy with a successful group positioned in a similar decision situation, groups could construct a viable decision rule for their own group.

Given the combination of manipulations (type of rule x way of inducement) we expect that: 1) the level of the group synergy in collaborative direct condition (CD) exceeds the group synergy in the collaborative analogical (CA) condition, and 2) that the level of group synergy in the identify-the-best direct (IBD) condition exceeds the level of group synergy in the IBA condition.

2.1    Methods & Results Study 1

146 students performed the Winter Survival exercise (Johnson & Johnson, 1987) first individually and then in groups. The task was to decide about the rank-order of 12 items from lowest to highest importance for their survival. In line with Larson (2007), weak cognitive synergy has been computed by subtracting the mean of individual scores in the group from the group score and strong cognitive synergy has been computed by subtracting the score of the best performing member of the group from the group performance score. In the current study we crossed two manipulations (decision rule and type of inducement), each with two possible conditions. We have used a between-group design In the direct inducement conditions, groups have been asked to employ either the method of group collaboration (CD) or the decision rule of identifying the best performing group member (IBD). In the analogical condition groups had to follow either the method of group collaboration or identify-the-best but this time they were not directly induced but while showing scenarios of successful groups following these decision rules.

Our results indicate that there are no significant differences between the collaborative and identify-the-best decision rule F (1,48)=0.09, p=0.75 for weak cognitive synergy nor for strong cognitive synergy F(1,48)= 0.28, p=0.59. There are also no differences between the two types of rule inducement, with F (1,48)=2.40, p=0.12 for weak cognitive synergy and F(1,48)=3.22, p=0.08 for strong cognitive synergy.

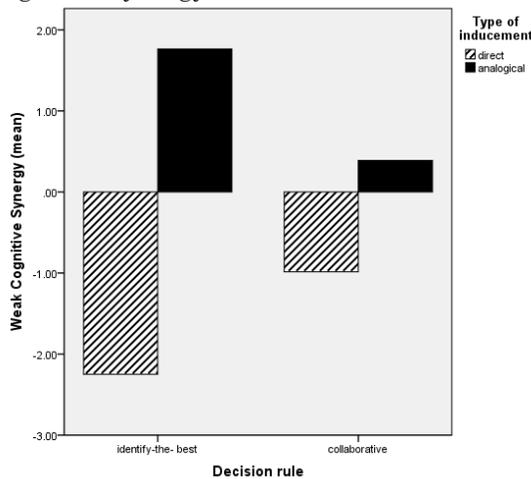 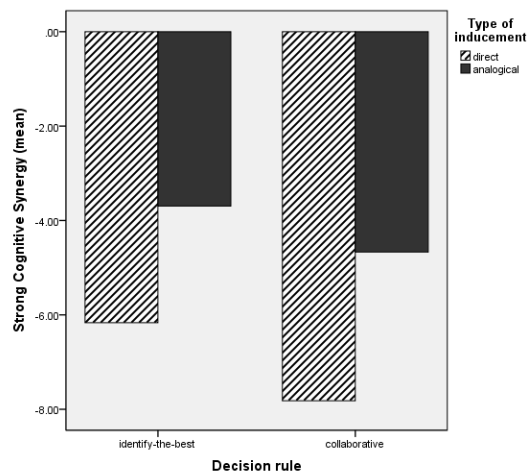

Figure 1             Figure 2

When looking at Figure 1 and 2 we further identify that contrary to our expectations, groups perform better in the analogical manipulation than in the direct manipulation, irrespective of the type of rule followed, for both weak and strong cognitive synergy. Interestingly, for weak cognitive synergy groups manage to reach absolute levels of synergy (scores are positive) only in the analogical manipulation, again irrespective of the type of rule followed. Our initial prediction was that groups following directly induced rules will outperform groups following analogical induced rules which involves an extra step in the process of establishing the group decision rules. One alternative explanation for this counterintuitive observation is that participants in the analogical conditions have more autonomy in defining their own decision rule, while groups with the direct rule manipulation have to follow an imposed decision rule. Choi and Levine (2004) indicate that groups that have a choice (high degree of autonomy) in defining their own working strategy are more committed to it and less prone to change it in a subsequent task.

## 3. STUDY 2

In order to clarify whether this alternative explanation is supported by our unexpected observations in Study 1, we have designed a second study in which we contrast four conditions. The first two conditions (self-selection) are the baseline conditions in which groups are allowed to decide their own rule: (1) uninformed self – selection: no decision rule, groups are free to select any decision rule and no further influence is being exerted on the groups (USS) and (2) informed self-selection: groups are free to develop their own decision rule with the ultimate goal of becoming better than their best performing group member (ISS). The last two conditions are induced decision rules selected from Study 1: CD and IBA. The goal of the second study is therefore to compare the two induced decision rule situations (CD and IBA) with the two self-selected conditions (ISS and USS). If the group's ability to reach cognitive synergy depends on the degree of autonomy in choosing a decision rule then the self-selection conditions should yield superior synergetic effects as compared to the induced decision rule.

### 3.1  Methods & Results Study 2

333 students had to perform the NASA task (Hall & Watson, 1970) first individually and then in groups (average size=4). The NASA task consists in deciding about the rank-order of 15 items from lowest to highest importance for their survival. Synergy scores have been computed similarly to Study 1. For strong cognitive synergy a significant mean difference has been identified between the USS (M=-2.47, SD=9.33) and IBA (M=3.05, SD=9.80), t= 6.51, p=0.02, CI [0.88; 12.13] as well as a significant difference between the ISS (M=-1.36, SD=8.87) and IBA, t= 7.10, p=0.03, CI [0.63; 13.56]. The comparison of conditions is also displayed in Figure 3 and 4.

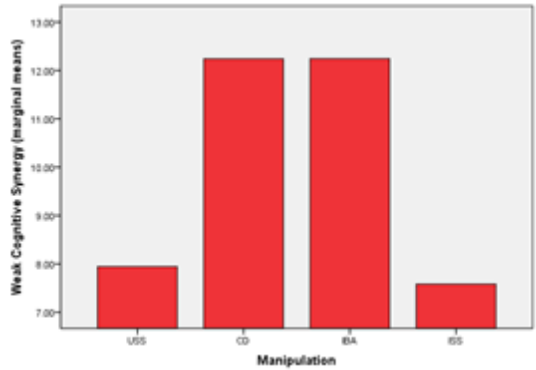
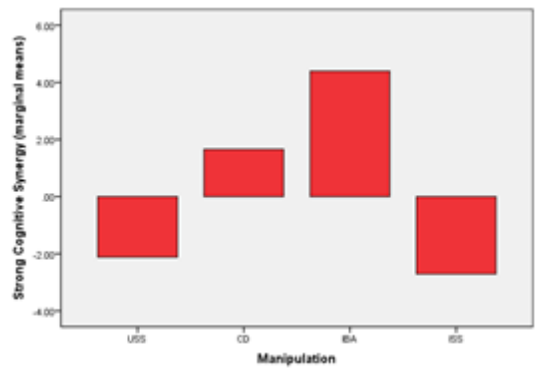

Figure 3                                    Figure 4

## 4. DISCUSSIONS AND CONCLUSIONS

Our paper contributes to the decision-making stream of research by indicating the beneficial effects of a heuristic decision rule (imitate-the-successful/ analogical inducement) on decision quality. This type of inducement proves to be a stronger manipulation than the content of the rule in itself. Practitioners should further consider not only the decision rule used to stimulate group synergy but also the way in which this decision rule is being communicated and induced. Secondly, we contribute to the cognitive synergy literature. Our findings indicate that strong group synergy is more likely to be achieved when groups (1) follow analogically induced decision rules rather than directly induced rules (2) follow the identify-the-best decision rule (induced analogically) rather than self-selected rules. This finding does have practical implications for group interventions.

---

[i] Paper accepted for publication in Plos One